\documentclass[prb,twocolumn,amsmath,floats,showpacs,superscriptaddress]{revtex4}

\usepackage{graphicx}% Include figure files
\usepackage{dcolumn}% Align table columns on decimal point

\begin{document}
\title{Formation and properties of metal-oxygen atomic chains}

\author{W.H.A. Thijssen}
\affiliation{Kamerlingh Onnes Laboratory, Leiden University, P.O.
Box 9504, 2300 RA Leiden, The Netherlands}

\author{M. Strange}
\affiliation{Center for Atomic-Scale Materials Physics, Department
of Physics, Technical University of Denmark, DK-2800 Lyngby,
Denmark}

\author{J.M.J. aan de Brugh}
\affiliation{Center for Atomic-Scale Materials Physics, Department
of Physics, Technical University of Denmark, DK-2800 Lyngby,
Denmark} \affiliation{Solid State Physics Group, MESA+ Research
Institute, University of Twente, P.O. Box 217, 7500 AE Enschede,
The Netherlands}

\author{J.M. van Ruitenbeek}
\affiliation{Kamerlingh Onnes Laboratory, Leiden University, P.O.
Box 9504, 2300 RA Leiden, The Netherlands}

\begin{abstract}
Suspended chains consisting of single noble metal and oxygen atoms
have been formed. We provide evidence that oxygen can react with
and be incorporated into metallic one-dimensional atomic chains.
Oxygen incorporation reinforces the linear bonds in the chain,
which facilitates the creation of longer atomic chains. The
mechanical and electrical properties of these diatomic chains have
been investigated by determining local vibration modes of the
chain and by measuring the dependence of the average
chain-conductance on the length of the chain. Additionally, we
have performed calculations that give insight in the physical
mechanism of the oxygen-induced strengthening of the linear bonds
and the conductance of the metal-oxygen chains.
\end{abstract}

\date{\today}

\maketitle
\section{introduction}
Freely suspended atomically thin metallic wires are the ultimate
one-dimensional conductors. Since their
discovery\cite{yanson98,ohnishi98}, scientists have been amazed by
the mere existence of these wires, since at first one would not
expect such one-dimensional structures to be able to physically
exist: The capillary instability that is also observed when
droplets break off a forming water column would destroy the
one-dimensionality since the surface tension is not strong enough
to keep the structure together. In fact atomic wires are only
observed when they are suspended between metal electrodes at both
sides or supported by a substrate onto which they can be
artificially be created atom by atoms \cite{nilius02,wallis02} or
formed by self-assembly \cite{gurlu03}. Once the wire breaks due
to a critical built-up of stress in the chain, the atoms collapse
back to the electrode since the necessary wire tension that kept
the structure together has disappeared. Atomic wires have actually
only been observed for the three 5d transition metals Au, Pt and
Ir in break-junction experiments at low temperatures
\cite{smit01}. Although there has been an indication for the
formation of short silver chains in TEM experiments at room
temperature in UHV \cite{rodrigues02}. It has been argued that the
physical mechanism that drives atomic chain formation is similar
to what drives spontaneous surface reconstructions in these same
metals, a delicate balance between the s and d electron density of
states influenced by relativistic effects \cite{takeuchi,pyykko}.
Due to the high charge of the nucleus the lowest s-electron
orbitals become relativistically contracted, which causes a
lowering of the Fermi energy resulting in a small depletion of the
anti-bonding states in the d-band. In the case of the 5d
transition metals this relativistic effect is large enough to tip
the energy balance for the formation of surface reconstructions
and the formation of atomic chains. This is not the case for 4d
and 3d transition metals \cite{smit01}. Indeed a Ag(110) surface
does form a missing row reconstruction while Au(110) does.

Experimentally freely suspended atomic wires can be created by
means of the mechanically controlled break junction (MCBJ)
technique \cite{muller92}, which is employed in this work. With
this technique a small junction can be mechanically broken and be
brought into contact again. Alternative approaches involve the use
of a STM, the tip of which can be contacted with a surface
\cite{agrait93} or by creating holes in a thin metallic foil by
electron bombardment until a single strand of gold atoms is left,
which can be imaged by TEM. \cite{ohnishi98,rodrigues01}. The MCBJ
technique and the electron bombardment technique have shown a
large discrepancy in the observed interatomic distances between
the atoms in the chain. For a gold chain created by the MCBJ
technique in cryogenic vacuum an interatomic distance of 2.5 $\pm$
0.2 \AA\ is observed \cite{untiedt02} while room temperature TEM
images give distances up to 4 \AA\ \cite{ohnishi98, rodrigues01}.
Based on model calculations, it has been suggested that these
large interatomic distances are due to the incorporation of
foreign atoms like oxygen \cite{bahn02}, hydrogen
\cite{skorodumova03}, carbon or nitrogen \cite{novaes03} that are
not resolved by TEM. Especially oxygen is an interesting candidate
since Density Functional Theory (DFT) calculations have shown that
the total energy of a gold atomic chain with atomic oxygen
incorporated is lowered compared to clean gold \cite{bahn02} and
the Au-O bond is able to sustain higher pulling forces before
breaking \cite{novaes06}. We have previously given experimental
evidence showing that oxygen can indeed be incorporated into
atomic gold chains and that oxygen induces chain formation for
silver, which is known not to form chains in pure form
\cite{thijssen06}. In this article we will provide further
evidence for oxygen incorporation by demonstrating the presence of
oxygen in the chain by local vibration-mode spectroscopy and by
calculations on the conductance properties of the chains, which
strongly support our experimental observations.

\section{experimental technique}
In our experiments we used poly-crystalline Au, Ag and Cu wires
with a diameter of 100 $\mu m$ and a purity of 99.999 $\%$. After
creating a circular notch in the middle of the wire, it is fixed
on a bendable substrate in a three point bending configuration. By
using a mechanical axle the wire can be broken and with a
piezoelectric element it is possible to control the distance
between the two electrodes with atomic precision. A detailed
description of the MBCJ technique can be found in
Ref~\onlinecite{agrait03}. Our experiments were performed in
cryogenic vacuum. Once the sample chamber is evacuated and cooled
down to 5~K the wire is broken for the first time ensuring clean
fracture surfaces. Close to the junction a local heater and
thermometer are mounted, which enable us to control the
temperature of the junction between 5 and 60~K, while the vacuum
can remains at 5~K. In order to be able to admit oxygen to the
junction the insert is equipped with a capillary attached to a
high purity oxygen reservoir. To prevent premature condensation of
oxygen, the capillary has a heating wire running all along the
interior. To prevent other molecules to enter into the sample
chamber while admitting oxygen through the heated capillary, it is
pumped and baked out by means of the heating wire at 150~$^{o}$C,
prior to the cool down. Two point conductance measurements are
performed to determine the conductance of the atomic wires.
Differential conductance (dI/dV) measurements were performed on
the atomic wires using a lock-in amplifier. The bias voltage was
modulated with a fixed modulation amplitude of 1 mV and a
frequency of 7 kHz, while sweeping the dc bias voltage from + 100
mV to - 100 mV and back.

\section{observation of oxygen induced atomic chains}
\begin{figure} [t!]
\includegraphics[width=8.5cm]{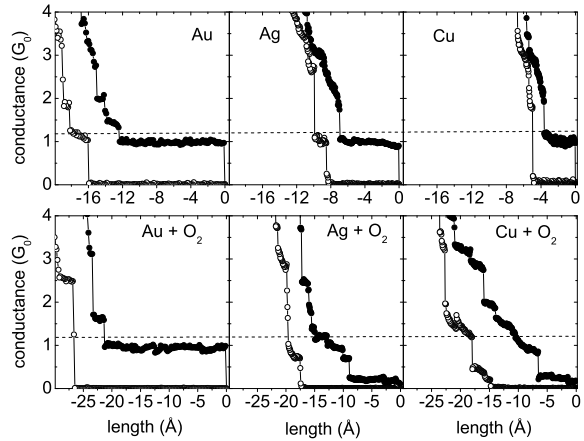}
\caption{\label{traces} Typical conductance traces for gold,
silver and copper with and without oxygen admitted. The filled
dots show the digitalized points measured while breaking the
contact, while the open dots are measured when closing the
junction again.}
\end{figure}

Microscopic junctions of gold, silver and copper are carefully
broken and the conductance is observed to decrease stepwise until
finally the contact has a cross section equal to the diameter of
only a single atom. For the noble metals the conductance of
single-atom contacts is close to one quantum unit of conductance,
G$_0$ = 2e$^{2}$/h. For gold it is known that upon stretching a
single atom contact, additional atoms can be pulled out of the
electrodes to form a monatomic chain of up to 7 atoms long
\cite{yanson98,hakkinen00,bahn01}. This manifests itself in long
plateaus with conductances around 1 G$_0$ that are observed while
monitoring the conductance of gold contacts as they are pulled
apart. Such long plateaus are not observed for silver and copper
as can be seen in Fig. \ref{traces}. This indicates that silver
and copper do not form long single atom wires like gold does.

\begin{figure} [t!]
\includegraphics[width=7cm]{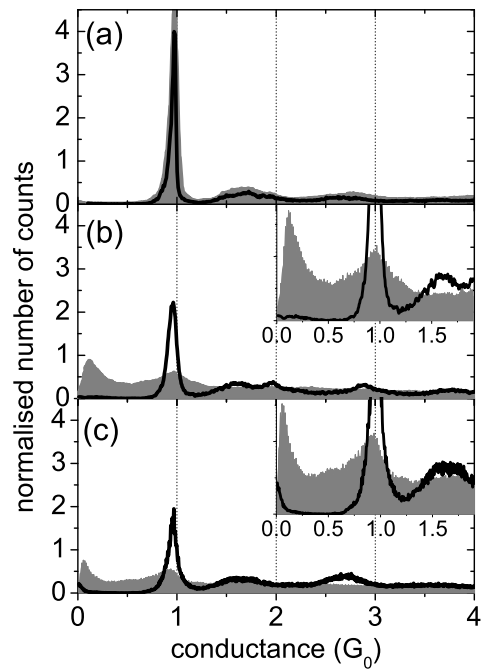}
\caption{\label{fig2} Conductance histograms for Au (a), Ag (b)
and Cu (c) without (black curves) and with oxygen admitted (filled
graphs). For clarity the histograms have been normalized to the
area under the curves. Each histogram is constructed from the
digitalized points (filled dots of Fig. \ref{traces}) obtained
from approximately 2000 curves recorded at a bias voltage of 50 mV
at T = 5 K.}
\end{figure}
When oxygen is admitted to the contact and the temperature is kept
at 5 K, interesting differences in the conductances and lengths of
the last plateaus are observed: In the case of gold the
conductance and the average length of the chains does not change
noticeably. This is in contrast to what is seen in the case of the
non-chain-forming metals silver and copper. The conductance of the
last plateau drops sharply towards 0.1 G$_0$ and 0.2 G$_0$ for
silver and copper, respectively, and the average total length
increases dramatically. Only when the temperature of the gold
sample is increased to about 40 K the average chain length also
increases. One way to obtain more quantitative information about
the conductance of the atomic contacts and chains is to record
conductance histograms. By recording all conductances that are
measured when a contact is being pulled apart and plotting them in
a histogram a conductance histogram is obtained. After breaking
and measuring many junctions the preferential conductances can be
determined. In Fig. \ref{fig2} conductance histograms for the
three noble metals are shown with and without oxygen admitted. In
order to focus on the changes that occur due to the admission of
oxygen the histograms have been normalized to the area under the
curves. For the clean metals one can clearly see that a peak near
1 G$_0$ is dominant over all other conductances. When oxygen is
introduced the dominant peak at 1 G$_0$ decreases sharply for
silver and copper due to a shift of weight to lower conductance
values. For gold at 5 K no significant changes are seen as was
discussed above. These observations indicate that oxygen
influences the atomic contacts of silver and copper in such a way
that configurations with conductances near 1 G$_0$ are less
frequently obtained. The conductance of atomic contacts for gold
is apparently hardly influenced by oxygen since the conductance
histograms of Fig. \ref{fig2} are nearly identical.

Recently we have presented evidence \cite{thijssen06} that Au
atomic wires become reinforced and therefore can be stretched to
longer lengths as a result of their reaction with oxygen at 40 K.
Furthermore, we have shown that this bond strengthening effect is
even more pronounced for silver wires. We have studied the chain
formation process under the influence of oxygen for the three
noble metals. Since the conductance of an atomic chain or
single-atom contact of noble metals are known to be close to 1
G$_0$ we can measure the lengths of a large ensemble of
conductance plateaus near 1 G$_{0}$ and construct a length
histogram. In such histogram the lengths of the plateaus are
plotted against the number of times those lengths were observed.
It can be clearly seen from the conductance histograms of Fig.
\ref{fig2} that in the case of silver and copper with oxygen
admitted a large portion of the observed conductances have values
below 1 G$_0$. We suspect that these low conductance values belong
to silver and copper atomic wires that are influenced by oxygen
and therefore we have included these conductances in the chain
length measurements. The length of the chains is obtained by
measuring the distance, which the electrodes move apart from the
moment when the conductance of the contact drops below 1.1 G$_0$
until it drops below 0.05 G$_0$. In the case of gold a conductance
limit of 0.5 G$_0$ was used for stopping the chain length
measurement.

\begin{figure} [t!]
\includegraphics[width=7cm]{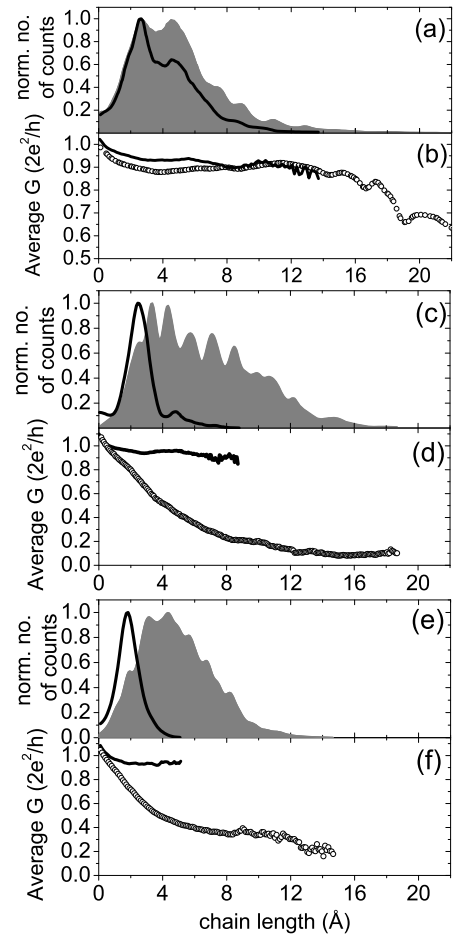}
\caption{\label{fig3} Length histograms of stretch distances for
contacts with conductance near 1 G$_{0}$ and lower of Au (a), Ag
(c) and Cu (e) in pure form (black curves) and with oxygen
admitted (filled graphs). For the pure metals the lengths were
measured in a conductance window of G $\in$ [1.1,0.5], while with
oxygen admitted G $\in$ [1.1,0.05]. The lower panels show the
average conductance as a function of stretch distance for clean Au
(b), Ag (d) and Cu (f) (black curves) and with oxygen admitted
(open dots). All data were obtained with a bias voltage of 50 mV.
The temperature graphs (a) and (b) was 40 K and for (c) to (f) 5
K.}
\end{figure}

In Fig. \ref{fig3} (a), (c) and (e) the length histograms for the
three clean noble metals and those with oxygen admitted are shown.
What is immediately seen is that the average chain length for all
three metals increases upon oxygen admission. This effect is
clearly the strongest for the non-chain-forming metals silver and
copper. In all length histograms one can clearly see a number of
more or less equidistant peaks. For silver-oxygen and
copper-oxygen chains the distances between the peaks are about 1.5
$\pm$ 0.2 \AA~ and 1.2 $\pm$ 0.2 \AA, respectively. These
equidistant peaks give a strong indication of atomic chains being
formed and the distances between the peaks have previously been
interpreted as the interatomic distance in the chains
\cite{yanson98}. When assuming that oxygen is indeed taken up in
the atomic chain structure, the interpretation of the peak
structure in terms of bond lengths for such a diatomic chain is
less straight forward. A bond length of only 1.2 \AA~is not
expected and is about 60$\%$ smaller than for the gold-oxygen
chains (see Fig.~\ref{fig3}a). DFT calculations have predicted
bond lengths of 1.9 \AA~ and 1.8 \AA~for silver-oxygen and
copper-oxygen chains respectively \cite{aandebrugh05}. The
shoulder to the left of the silver-oxygen length histogram of Fig.
\ref{fig3} is located at the same position as the dominant peak in
the clean silver histogram. Therefore it is most probably caused
by the rupture of a $>$Ag-Ag$<$ contact. Here we have introduced
the notation $>$ and $<$ that indicate the connections to the left
and right electrodes, respectively. We will take a more detailed
look at the peak positions that arise in the length histograms of
Fig.~\ref{fig3}(c), in order to explain the discrepancy. A length
count is taken at the moment an atomic chain breaks and the atoms
collapse back to the electrodes, which causes the the conductance
of the junction to drop deep into the tunneling regime. A rupture
can occur when one of the linear one-dimensional bonds cannot be
stretched any further due to a critical built-up of energy in the
chain. But sometimes it is possible that an additional atom, which
is weakly bound to the bulk atoms at one of the electrode apexes,
is pulled into the atomic chain. Consequently, the chain remains
intact and the tension of the linear bonds is relaxed. Then it is
possible to stretch the one-dimensional structure further until
again a critical built-up of energy occurs, which can result into
rupture or the elongation of the chain with yet another atom. This
sequence repeats itself over and over again and the incorporation
of atoms will stop when the supply of loosely bound atoms at the
apexes of the electrodes is depleted. The peaks in the length
histograms indicate lengths at which there is a higher preference
for the atomic chain to break and therefore represent atomic chain
lengths in a stretched configuration.

In table~\ref{table} all peak positions together with an
interpretation in terms of the corresponding chain composition are
listed. From the length histogram of clean silver in
Fig.~\ref{fig3} we obtain a silver-silver bond length of 2.4 \AA\
and in order to simplify the analysis we kept this value fixed.
Furthermore we assume that the bond length at the edges where the
atomic chain is contacted to the electrodes with either a silver
($>$Ag) or an oxygen atom ($>$O) is the same. From this simple
analysis we arrive at a silver-oxygen bond length of 1.8 $\pm$ 0.1
\AA, very close to the theoretical value
\cite{aandebrugh05,dasilva06}. A similar analysis can be made for
the peak positions of the copper-oxygen length histogram of Fig.
\ref{fig3}(e), which yields a copper-oxygen bond of 1.7 $\pm$ 0.1
\AA.

\begin{figure} [t!]
\includegraphics[width=9cm]{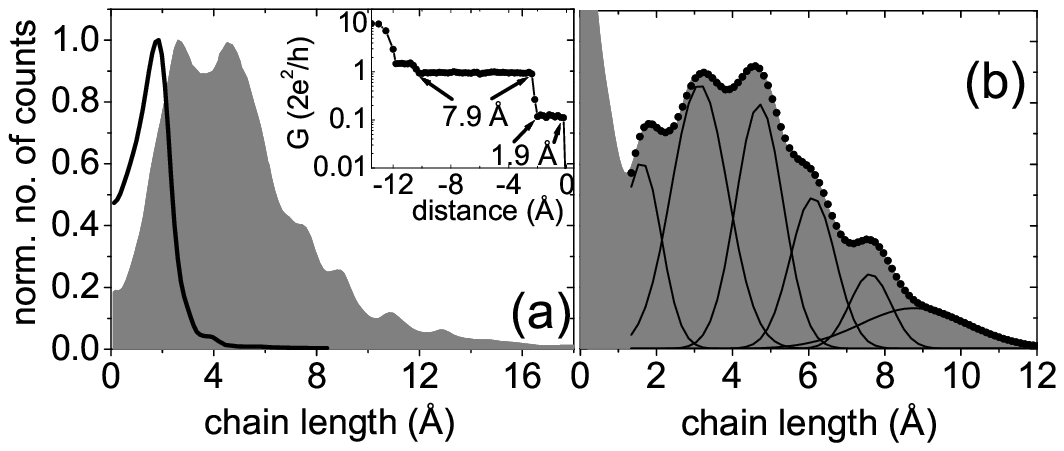}
\caption{\label{fig4} (a) Length histogram for Au-O chains at 40 K
with conductance limits between 1.1 and 0.5 G$_{0}$ (filled graph)
compared to a length histogram with conductance limits between 0.2
and 0.05 G$_{0}$ (black curve); (b) Length histogram of Ag-O
chains at 5 K with conductance values between 0.5 and 0.05
G$_{0}$. All histogram were taken at a bias voltage of 50 mV and
consist of at least 2000 traces.}
\end{figure}

The decrease of the conductance of the chains as the chains become
longer is very nicely seen in Fig.~\ref{fig3} (b), (d) and (f).
While for silver-oxygen and copper-oxygen chains the drop in
conductance starts already for relatively short chains,
gold-oxygen chains display a lower conductance only when they are
on average longer than about 16 \AA, corresponding to a chain with
a length of more than six gold atoms.
\begin{table}[b]
\begin{center}
\begin{tabular}{|p{1.2cm}|l|p{1.2cm}|p{1.2cm}|}
 \hline
  \small{peak position (\AA)} & \small{proposed chain compositions} & \small{number of Ag-O
  bonds} & \small{length of Ag-O bond} \\ \hline
  3.3& $>$Ag-O-Ag$<$ or $>$O-Ag-O$<$& 2 & 1.7 \AA \\ \hline
  4.3 & $>$Ag-Ag-O$<$ & 1 & 1.9 \AA \\ \hline
  5.7 & $>$Ag-Ag-O-Ag$<$ & 2 & 1.7 \AA \\
   & $>$Ag-O-Ag-O$<$ & 3 & 1.9 \AA \\ \hline
  7.1 & $>$Ag-O-Ag-O-Ag$<$ & 4 & 1.8 \AA \\ \hline
  8.5 & $>$Ag-Ag-O-Ag-Ag$<$ & 2 & 1.8 \AA \\
   & $>$Ag-O-Ag-O-Ag-O$<$ & 5 & 1.7 \AA \\ \hline
  9.9 & $>$Ag-Ag-O-Ag-Ag-O$<$ & 3 & 1.7 \AA \\
   & $>$Ag-Ag-O-Ag-O-Ag$<$ & 4 & 1.9 \AA \\ \hline
  10.7 & $>$Ag-O-Ag-O-Ag-O-Ag$<$ & 6 & 1.8 \AA \\ \hline
  12.2 & $>$Ag-Ag-O-Ag-O-Ag-Ag$<$ & 4 & 1.9 \AA \\
   & $>$Ag-O-Ag-O-Ag-O-Ag-O$<$ & 7 & 1.7 \AA \\ \hline
  14.8 & $>$Ag-Ag-O-Ag-O-Ag-O-Ag-Ag$<$ & 6 & 1.7 \AA \\
   & $>$Ag-O-Ag-O-Ag-O-Ag-O-Ag$<$ & 8 & 1.9 \AA \\
  \hline
\end{tabular}
\end{center}
\caption{\small{Silver-oxygen chain configurations that fit the
experimental peaks positions from the length histogram of
figure~\ref{fig3}(c). The numbers and lengths of the silver-oxygen
bonds are also given.}} \label{table}
\end{table}
Indeed for Au-O chains created at 40 K it is sometimes observed
that upon stretching the conductance makes as sudden jump from a
value around 1 G$_{0}$ to about 0.1 G$_{0}$ as can be seen in the
inset of Fig.~\ref{fig4}(a). We have investigated these lower
conduction plateaus by recording a length histogram between
conductance values of 0.2 and 0.05 G$_{0}$. In
figure~\ref{fig4}(a) the length histogram for low conductance
values of Au-O chains is shown. It is dominated by a single peak
at 1.9 \AA. A second small peak can be seen at 3.9 \AA. This
distance is the same as the Au-O distance that we observed in the
length histogram of Fig.~\ref{fig3}(a). Furthermore it is striking
that plateaus at 0.1 $G_{0}$ are typically only one bond length
long and only occur when a chain of certain length has already
been formed.
We conclude, therefore, that the atomic unit that is being pulled
into the chain and causes the sharp drop in conductance, also
destabilizes the chain since upon further pulling the chain breaks
in nearly all cases. The plateaus at 0.1 G$_{0}$ have only been
seen at temperatures $\geq$ 40 K. The higher mobility and vapor
pressure of oxygen molecules at those temperatures increases the
supply of oxygen molecules around the region where the chain is
formed. We speculate that the low conductance plateau is caused by
the appearance of oxygen-oxygen bonds in the chain which could
severely lower both the conductance and the bond strength.

A length histogram obtained for low-conductance silver-oxygen
chain structures is shown in Fig.~\ref{fig4}(b). In contrast to
gold (Fig.~\ref{fig4}(a)) chain formation continues by
incorporating more atomic units into the chain since multiple
peaks in the length histogram are observed. The inter-peak
distance is 1.5 $\pm$ 0.2 \AA, which is the same as in the
histogram of Fig.~\ref{fig3}(c), indicating that both silver and
oxygen atoms are being pulled into the chain.

\begin{figure} [b!]
\includegraphics[width=6cm]{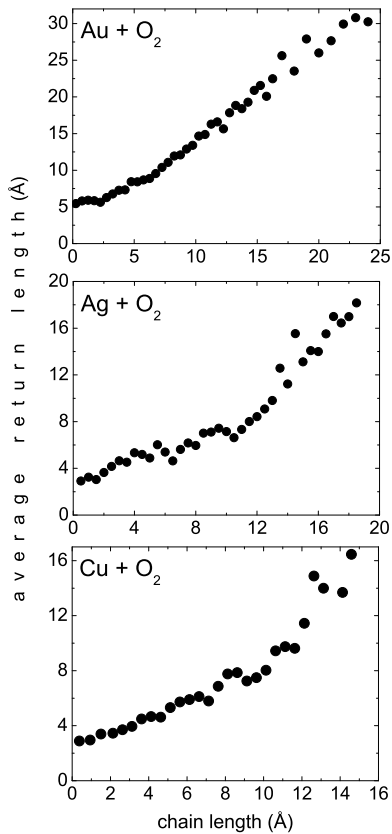}
\caption{\label{fig5} Average return lengths as a function of
atomic chain length for Au (a), Ag (b) and Cu (c) with admitted
oxygen at respectively 40 K, 5 K and 5 K.}
\end{figure}

More evidence to support that freely suspended long atomic wires
are being formed of gold, silver and copper with chemisorbed
oxygen incorporated comes from the dependence of the average
return length on the chain length. When an atomic wire is formed,
atoms are pulled out of the electrodes in order to be incorporated
into the atomic wire upon stretching the contact. When a wire
breaks, the atoms collapse back to the electrodes and the
conductance drops deep into the vacuum tunneling regime. We then
reverse the direction of movement of the electrodes towards each
other. At the moment that mechanical contact is remade the
conductance jumps from a value in the tunneling regime to a value
of order 1 G$_0$. By measuring this so-called return length for
every wire that is broken we obtain the average return length for
a given atomic chain length. Fig.~\ref{fig5} shows the average
return length for gold-oxygen, silver-oxygen and copper-oxygen
chains. In all cases an offset is observed followed by an increase
in average return length as a function of chain length. The offset
indicates that even for a single atom contact that breaks before
forming a chain the electrodes move back a certain distance. This
is due to the fact that the single atom connects the two
electrodes and it can sustain a force upon pulling on it. This
causes a counter-force pulling on the electrodes, which will
slightly elastically deform them. When the contact breaks the
stress can be released and the atoms relax and move back into
their equilibrium position. But when longer chains are being
formed and eventually broken the atoms that made up the chain
collapse back to the electrodes and consequently the electrodes
have to move back a longer distance upon increasing chain length.
Given the fact that the average return length increases more or
less linearly with chain length we have a strong indication that
atomic chains of noble metal atoms with oxygen are indeed being
formed. The deviation from a purely linear increase can be the
result of a larger energy stored in the longer chains, because
more metal oxygen bonds are present that can sustain higher
pulling forces. When such long chains break the stored elastic
energy may propel atoms away from the junction area. The force a
silver-oxygen bond in an infinite alternating silver-oxygen chain
can sustain is about 2.4 times larger than for a silver-silver
bond in a infinite silver chain \cite{aandebrugh05}. While the
difference in the case of pure gold and gold-oxygen is about a
factor 1.5 \cite{bahn02,novaes06} explaining why this effect is
not appearing so clearly in gold-oxygen return lengths.

\section{point contact spectroscopy on metal-oxygen chains}

In the previous section we have provided evidence that oxygen can
chemically react with noble metal contacts or chains and reinforce
the linear bonds in the wires so that they can be elongated
further. In this section we present evidence that oxygen most
probably is dissociated and incorporated into the atomic chains as
is predicted by DFT model calculations \cite{bahn02,novaes06}.

\begin{figure} [t!]
\includegraphics[width=7cm]{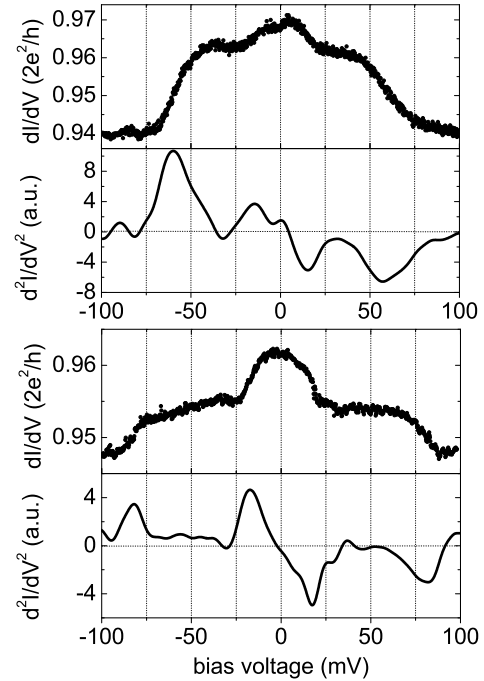}
\caption{\label{fig6} Two typical differential conductance spectra
taken for Au atomic chains after admitting oxygen at 4.2 K. The
chains were pulled to a length of about 7 \AA}
\end{figure}

We have verified the incorporation of oxygen by point contact
spectroscopy. This technique forms a powerful tool for studying
the interactions of electrons with vibrational excitations in a
metallic contact \cite{yanson86,khotkevich89}. An atomic chain of
noble metal atoms is a ballistic conductor for low bias voltages,
since it has one nearly completely transparent conductance channel
resulting in a total conductance very close to 1 G$_0$
\cite{scheer98,brom99,scheer01}. However, electrons with
sufficient energy have a small probability to excite phonon modes
in an atomic chain and by means of differential conductance
(dI/dV) spectroscopy the energies for those modes can be
determined as was shown for clean gold atomic wires
\cite{agrait02}. About 1~$\%$ of all forward traveling electrons
with an excess energy larger than the energy of a phonon mode
actually do excite such mode. Those electrons lose energy, which
forces them to scatter back since all forward moving states below
E$_{F}$ + eV$_{bias}$ are already occupied for a perfect
single-channel conductor. This backscattering causes a small
decrease of the total conductance of the atomic chain. The
observation of this small signal is often hampered by conductance
fluctuations, which result from interference of electron
trajectories scattering off defects near the contact
\cite{ludoph99,ludoph00}. In the case of the s-like noble metals
that have a nearly 100$\%$ transparent single channel these
conductance fluctuations are strongly suppressed \cite{ludoph99}.
From Fig.~\ref{fig3} we see that gold-oxygen chains have a
conductance that is close to 1 G$_0$ and we have observed phonon
modes for medium-length gold-oxygen wires. Differential
conductance spectra were recorded by keeping the atomic chain at a
fixed elongation and measuring the current modulation as was
described before. Fig.~\ref{fig6} displays two spectra that were
obtained for 7 \AA~long gold-oxygen wires at 5 K. Clearly two
downward steps at different voltages in each spectrum can be seen,
indicating a possible onset of phonon excitations at those
energies. The low-energy phonon mode has an energy at 17 $\pm$ 2
meV in Fig.~\ref{fig6}(a) and 15 $\pm$ 2 meV in
Fig.~\ref{fig6}(b). The high energy mode that is observed has an
energy of 82 $\pm$ 4 meV and 60 $\pm$ 4 meV in Fig.~\ref{fig6}(a)
and Fig.~\ref{fig6}(b) respectively. All spectra that we have
observed, which display two phonon steps, have high energy modes
between 60 and 80 meV.

In order to give a qualitative explanation of the observed
vibration mode energies, let us take a look at a simple classical
spring-mass model. The model consists of two gold atoms attached
to solid walls and an oxygen atom connected by springs in between
the gold atoms. We consider only the longitudinal
eigenfrequencies, because the probability of exciting transversal
modes is very small given the quasi one-dimensional geometry of
the atomic chain. By solving the equations of motion we obtain
three frequencies for masses {\it M} and {\it m} for gold and
oxygen, respectively. Here we have used similar spring constants
$k$ between all masses for simplicity, and obtain
\begin{eqnarray}\label{model}
\begin{array}{cc}
\omega_{1}^{2} = \frac{M + m - \sqrt{M^{2} + m^{2}}}{Mm}k \\
\\
\omega_{2}^{2} = \frac{2}{M}k \\
\\
\omega_{3}^{2} = \frac{M + m + \sqrt{M^{2} + m^{2}}}{Mm}k \\
\end{array}
\end{eqnarray}

The lowest frequency mode $\omega_{1}$ corresponds to the in-phase
motion of all atoms. In the second mode $\omega_{2}$ the oxygen
atom is immobile and the gold atoms move in anti-phase. The
highest frequency mode $\omega_{3}$ corresponds to the oxygen atom
moving in anti-phase with the gold atoms. By substituting the
masses for gold and oxygen atoms, we find that the ratio of two
low energy modes $\omega_{1}$ and $\omega_{2}$ is 1.4. The
heavy-mass modes will be close to the energy of the longitudinal
mode for clean atomic gold chains, that are found at about 10-15
meV depending on stretching\cite{agrait02}. From Eq.(\ref{model})
the high-energy mode can then be expected at 65 $\pm$ 10 meV.
Preliminary DFT calculations on gold-oxygen chains predict a high
energy mode in the range of 60 to 80 meV \cite{dasilva06},
depending on the stress in the linear bonds. Fig.~\ref{fig6} shows
a high energy mode at 60 meV and one at 80 meV, comfortably in the
range of our simple model and the preliminary calculations.
Additionally, our analysis of the gold-oxygen vibration modes
provides evidence that oxygen atoms, not molecules, are
incorporated in the atomic chain: An oxygen molecule has double
the mass of an atom, which would shift the energy of the vibration
mode down by a factor $\sqrt{2}$ to 46 $\pm$ 7 meV, assuming a
similar bond strength to gold, clearly off the experimental
values.
\begin{figure} [t!]
\includegraphics[width=7cm]{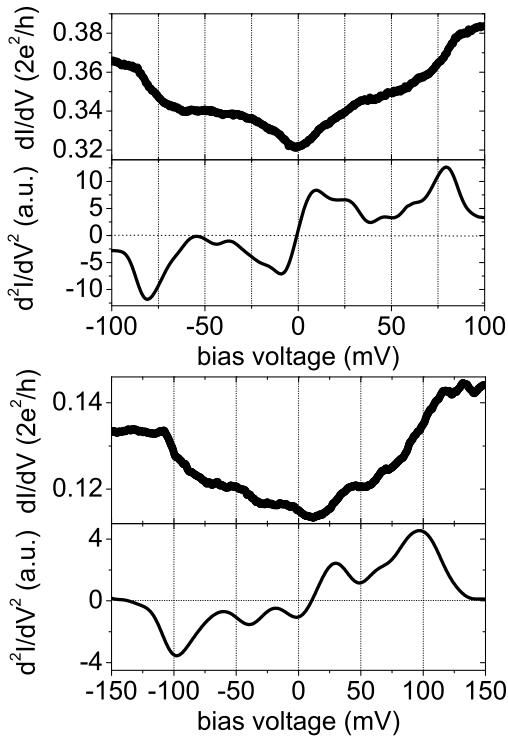}
\caption{\label{fig7} Two $dI/dV$ spectra taken for Ag atomic
chains after admitting oxygen at 5 K. The top spectrum was
obtained for a chain of about 7~\AA~in length and the bottom graph
for a 11~\AA~long chain. The $d^{2}I/dV^{2}$ spectra have been
smoothened for clarity.}
\end{figure}

We also performed point contact spectroscopy on silver-oxygen
atomic chains. In contrast to the gold-oxygen atomic chains, the
conductance of the silver-oxygen chains is typically much lower
than 1 G$_{0}$, which indicates that the conductance is not made
up from a single fully transparent conduction channel. Normally,
when one deals with one or more partly opened conduction channels
the dI/dV spectrum is dominated by conductance
fluctuations\cite{ludoph00}. This makes the observation of steps
in dI/dV due to inelastic scattering of electrons on local
vibration-modes very difficult. Furthermore it was recently
claimed that the inelastic correction to the conductance changes
from a negative contribution for a single conduction channel with
G $>$ 0.5 G$_{0}$ to positive for G $<$ 0.5 G$_{0}$
\cite{viljas05,paulsson05,vega06}. We obtained a few spectra for
silver-oxygen chains with a conductance $\leq$ 0.5 G$_{0}$, which
displayed bias-symmetric step-like features at energies far above
20 meV (see Fig.~\ref{fig7}).

One can distinguish clearly a step-like increase of the
conductance around 80 and 100 meV for the top and bottom spectra,
respectively, which would indicate forward scattered inelastic
electrons, in agreement with theory \cite{vega06}. The energies of
80-100 meV are in the range where local vibration modes of these
one-dimensional structures can be expected, albeit that compared
to the 60-80 meV for the gold-oxygen chains they are somewhat
high. Since the energy is far above the Debye energy for silver,
the modes are related to incorporated oxygen in the chain.

\section{calculations}
\begin{figure} [b!]
\includegraphics[width=7 cm]{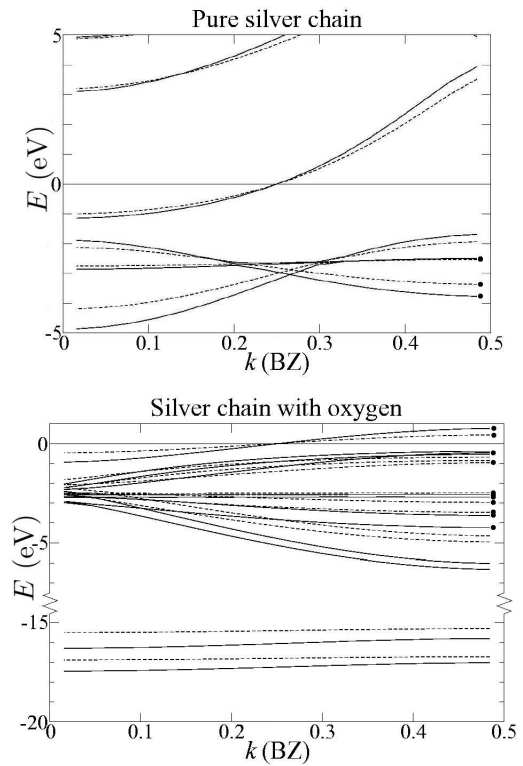}
\caption{Upper graph: Band structure for an infinite pure silver
chain with an interatomic distance of 2.6\,\AA\ (continuous lines)
and 2.8\,\AA\ (dashed lines). No spin polarization. Lower graph:
Band structure of an infinite alternating silver-oxygen chain with
an interatomic distance of 2.0\,\AA\ (continuous lines) and
2.2\,\AA\ (dashed lines). Bands are split in spin up and spin down
bands and degenerate bands are marked with a dot. In both cases,
32 $k$-points are used for the whole Brillouin zone.}
\label{AgBands}
\end{figure}

In this section we present DFT calculations for the conductance,
bonding properties and vibrational properites of Ag-O contacts.
Electronic structure calculations are performed using a plane wave
implementation of DFT.\cite{ksdft} Exchange and correlation
effects are treated at the GGA level using the PW91 energy
functional\cite{pw91} and the nuclei and core electrons are
described by ultrasoft pseudopotentials.\cite{ultrasoft} The Kohn
Sham (KS) eigenstates are expanded in plane waves with a kinetic
energy less than 400~eV. The width of the Fermi-Dirac distribution
for occupation numbers is set to 0.1~eV and the total energies are
extrapolated to T = 0.

We first consider a linear Ag chain with an interatomic distance
of $2.6~\text{\AA}$ and a linear alternating Ag-O chain with an
interatomic distance of $2.0~\text{\AA}$. The supercell for the
chains has transverse dimensions $12\text{\AA}\times12~\text{\AA}$
and we align the chains to the z-direction.

The band structure for Ag and Ag-O chains are shown in Fig.
\ref{AgBands} in the upper and lower graph, respectively. The
dashed lines indicate slightly strained chains, with the
interatomic distance increased by $0.2~\text{\AA}$.

The Ag chain is not magnetic and has fully occupied 4d bands and a
half filled 5s band. For the Ag-O chain we find a magnetic moment
of $1.0~\mu_B$ per Ag-O atom unit and the energy gain, due to the
spin-polarization is $0.12~\text{eV}$ per Ag-O atom unit.

The bands for the Ag-O chain can be grouped according to the
angular momentum quanta, m, in the chain direction. The two lowest
lying bands are the spin-split $\text{O}~2s$ (m=0) bands. The
remaining bands shown are $\text{Ag}~4d$ bands ($m=\pm2$) and
$\text{Ag}~4d-\text{O}~2p$ hybrid bands ($m=0,\pm1$). Straining
the Ag-O chain changes the profile of all bands (dashed lines).
However, only two of the spin bands with $d-p$ character change
considerably indicating that these bands are relevant for the
covalent bonding between Ag and O. The two bands are singly
degenerate since they have m=0 and corresponds to a $d_{z^2}-p_z$
hybrid band with bonding character. Novaes {\it et.
al}\cite{novaes06} recently showed that the covalent bond between
Au and O is mainly due to the bonding state between
$\text{Au}~5d_{z^2}$ and $\text{O}~2p_z$, in agreement with our
findings. It has been argued before that an interesting connection
exists between the observation of surface reconstructions and the
atomic chain formation in the metals gold, platinum and
iridium.\cite{smit01} In this context it is remarkable that oxygen
induced surface reconstructions on Ag(110) and Cu(110) have been
observed,\cite{tanaguchi92,besenbacher93} while we here provide
experimental evidence for atomic chain chain formation for these
metals upon oxygen chemisorption. Therefore we suspect that the
oxygen chemisorption on metal surfaces and the consequent
reconstruction originates from the $d-p$ hybrid bands playing the
same role as the d-bands in the chain forming metals.

There is a doubly degenerate spin band crossing the Fermi level in
Fig. \ref{AgBands} with $m=\pm1$, i.e. a $d_{xy}-p_x$ band and a
$d_{yz}-p_y$ band. This shows, that magnetic Ag-O chains are
metallic and can support two spin channels for a total maximum
conductance of $1~\mathrm{G_0}$. However, our experiments show,
that finite Ag-O chains often display conductances considerably
below $1~\mathrm{G_0}$ conductances.

To investigate this, we have calculated the spin-paired
transmission function for finite Ag-O wires with semi-infinite
silver chains as leads. For comparison we have also performed
calculations with gold instead of silver. We use a Green's
function method for phase coherent electron transport, where both
the Green's function of the semi-infinite leads and the scattering
region are evaluated in terms of basis consisting of maximally
localized Wannier functions.\cite{thygesen}

\begin{figure}[h!]
\includegraphics[width=7cm]{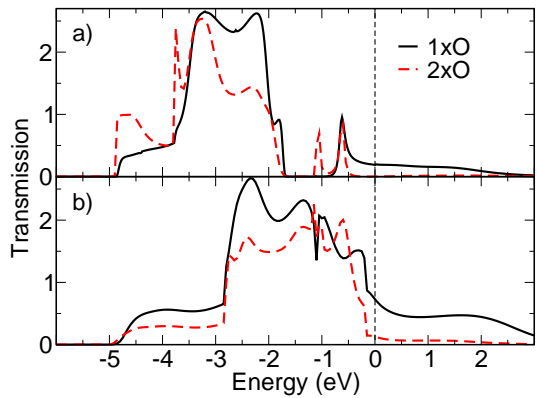}
\caption{(a) Transmission function for a $>$Ag-O$<$ (full curve)
and $>$Ag-O-Ag-O$<$ (dashed curve) wire contacted by semi-infinite
silver atomic wires. (b) Transmission function for a
$>$Au-O$<$(full curve) and $>$Au-O-Au-O$<$ (dashed curve) wire
contacted by semi-infinite gold atomic wires.}
\label{fig.TransmissionAgOAuO}
\end{figure}

Fig. \ref{fig.TransmissionAgOAuO} (a) and (b), show the
transmission functions for finite Ag-O and Au-O wires sandwiched
between Ag and Au chains, respectively. The full line is for a
$>$Ag/Au-O$<$ wire sandwiched between leads and the dashed line is
for $>$Ag/Au-O-Ag/Au-O$<$ wire between the leads. The Fermi level
is at zero and indicated by the vertical dashed line.

The transmission function for $>$Ag-O$<$ displays one eigenchannel
in the energy range $-2~\text{eV}$ to $3~\text{eV}$, indicating
that the eigenchannel state has $m=0$. The only available state on
the O atom with $m=0$ in the relevant energy range is the $2p_z$,
which is then the current carrying state on the O atom. This is
confirmed by the projected density of states for $\text{O}~2p_z$
(not shown), showing the resonance in the transmission function
below the Fermi level originating from an anti-bonding combination
of the $\text{O}~2p_z$ orbital and the $\text{Ag}~d_{z^2}$ band in
the leads. The broadening of the resonance is due to the coupling
to the s-band in Ag. The bonding combination falls outside the Ag
lead bands and is a bound state. In the case of $>$Ag-O-Ag-O$<$
units, the two resonances have a similar origin, but the
transmission at the Fermi level is reduced considerably.

The transmission function for the Au-O system shown in Fig.
\ref{fig.TransmissionAgOAuO} (b) consists of single eigenchannel
with m=0, in the energy range from -0.15~eV to $3.0~\text{eV}$.
Below $-0.15~\text{eV}$, there are Au d-bands with $m=\pm1$ which
can couple to O $m=\pm1$ states. This results in two extra
eigenchannels making the $2p_z$ resonance less visible accept for
the tail starting at $-0.15~\text{eV}$ and extending to
$3~\text{eV}$.

As seen in Fig.\ref{fig.TransmissionAgOAuO} (a) and (b) an oxygen
atom scatters more strongly electrons around the Fermi level in
the $m=0$ channel in the case of Ag than for Au. This in agreement
with our experiments, as we observed in Fig.\ref{fig3} a tendency
for Ag-O chains to have a lower conductance than Au-O chains.
However, in a more realistic contact with surface electrodes as
leads, we suspect a finite coupling between $m=\pm1$ finite $d-2p$
hybrid chain states and the $d$ states in the metal electrodes,
which would result in a higher conductance. A conductance
calculation on Ag-O chains involving more realistic electrodes
will be published elsewhere.

We also studied the longitudinal vibration-mode energies of a
single oxygen atom in a silver contact, we define a supercell with
$3\times3$ atoms in the surface plane and which contains 7 atomic
layers. The oxygen atom is clamped between two 4 atom pyramids
pointing in opposite directions and is attached to Ag (111)
surfaces. The Ag (111) surfaces are separated by
$12.7~\text{\AA}$. We use a $4\times4$ k-point Monkhorstpack mesh
in the surface Brillouin zone. The two pyramids and the oxygen
atom, were relaxed, until the total residual force was below
$0.05~\text{eV/\AA}$. The calculated longitudinal vibrational mode
energies for the two apex silver atoms and the oxygen atom are:
$\omega_1=81.4~\text{meV}$, $\omega_2=28.5~\text{meV}$ and
$\omega_3=14.7~\text{meV}$. The 3 vibrational modes corresponds
to: 1) The oxygen atom moving in anti-phase with the silver atoms.
2) The two Ag atoms moving in anti-phase and the oxygen atom being
immobile. 3) In phase motion of all 3 atoms.

The relation between the vibration-mode energies can be accounted
for by a simple model as described in the previous section and we
indeed see that the value of the calculated high energy mode is in
good agreement, with the measured vibrational energies in the
range $(80-100)~\text{meV}$, indicating that single oxygen atoms
are indeed incorporated in the chains.

Recently Ishida \cite{ishida07} presented calculations for chains
of Au and Ag atoms with a single O atom inserted. Ishida's results
also show a high transmission at the Fermi energy for oxygen in Au
chains, although for a single O atom it has more the character of
a resonance. For Ag the conductance is more strongly suppressed.

\section{conclusion}
We have studied the effects of oxygen on the atomic chain
formation for the noble metals gold, silver and copper. By
studying the conductance and mechanical properties of these atomic
chains, we provided evidence that oxygen atoms can be incorporated
in atomic chains. While atomic gold chains retain a conductance
near 1 G$_{0}$ when oxygen is incorporated, the conductance is
much reduced for silver and copper chains that form upon
chemisorption of oxygen atoms. The increased linear bond strength
makes it possible to form silver-oxygen and copper-oxygen atomic
chains. Our analysis of the interatomic distances in the
silver-oxygen and copper-oxygen chains has indicated the presence
of metal-oxygen bonds in the chain. The point contact spectra
obtained on the metal-oxygen chains give additional evidence that
atomic (not molecular) oxygen is incorporated in the atomic
chains. The observation that oxygen can be incorporated in atomic
chains and thereby increasing the linear bond strength could open
possibilities for other molecules to be incorporated in atomic
chains. This could lead to interesting new physical properties of
these ultimate one-dimensional structures.

This work is part of the research program of the ``Stichting
FOM,'', which is financially supported by NWO.

\end{document}